\documentclass[aps,pra,twocolumn,superscriptaddress,groupedaddress]{revtex4-1}  

\usepackage{graphicx}  
\usepackage{dcolumn}   
\usepackage{bm}        
\usepackage{amssymb}   
\usepackage{amsmath}
\usepackage{graphicx}
\usepackage{float}

\begin{document}

\title{Modeling quantum cascade lasers: Coupled electron and phonon transport  far from equilibrium and across disparate spatial scales}

\author{Y. B. Shi}
\author{S. Mei}
\author{O. Jonasson}
\author{I. Knezevic}\email{iknezevic@wisc.edu}

\affiliation{Department of Electrical and Computer Engineering, University of Wisconsin-Madison, Madison,
WI 53706, USA}

\begin{abstract}
Quantum cascade lasers (QCLs) are high-power coherent light sources in the
midinfrared and terahertz parts of the electromagnetic spectrum. They are devices in which the electronic and lattice systems are far from equilibrium, strongly coupled to one another, and the problem bridges disparate spatial scales. We present our ongoing work on the multiphysics and multiscale simulation of far-from-equilibrium transport of charge and heat in midinfrared QCLs.
\end{abstract}

\maketitle


\section{Introduction}

Quantum cascade lasers (QCLs) are electrically driven, unipolar, coherent light sources in the midinfrared (mid-IR) and terahertz (THz) parts of the electromagnetic spectrum \cite{Faist1994,Capasso2000}. In addition to being of great technological importance, QCLs are fascinating nonequilibrium systems that are typically thoroughly characterized via electrical, optical, and thermal measurements
precisely because of their practical value \cite{Yao2012,Williams2007,Vitiello2015}. As a result, QCLs are excellent as model systems for far-from-equilibrium theoretical studies \cite{Wacker2002a,bugajski2014mid}.

\begin{figure}[h!]
\centering
  \includegraphics[width=1.6 in]{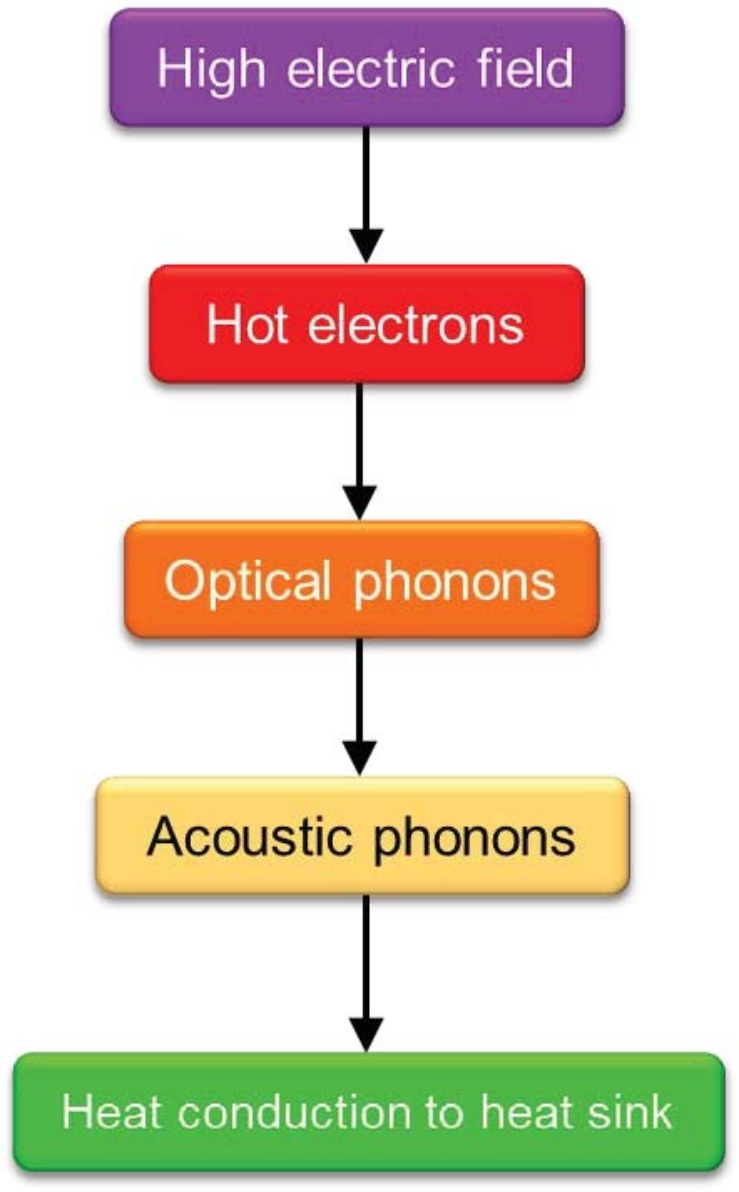}%
  \caption{\label{fig:energyflow}
    Flow of energy in a quantum cascade laser.}
\end{figure}

Under high-power, continuous-wave (CW) operation, the electron and phonon systems in QCLs are both very far from equilibrium and strongly coupled to one another, which makes them very challenging to accurately model. The problem of their coupled dynamics is both multiphysics (coupled electronic and thermal) and multiscale (bridging between a single stage and device level). During typical QCL operation, large amounts of energy are pumped into the electronic system, of which a small fraction is given back through the desired optical transitions, while the bulk of it is deposited into the optical-phonon system. Longitudinal optical (LO) phonons decay into longitudinal acoustic (LA) phonons; this  three-phonon process is often referred to as anharmonic decay. LA phonons have high group velocity and are the dominant carriers of heat. Figure \ref{fig:energyflow} depicts a typical energy flow in a QCL.

\begin{figure}[h!]
  \centering
  \includegraphics[width=\columnwidth]{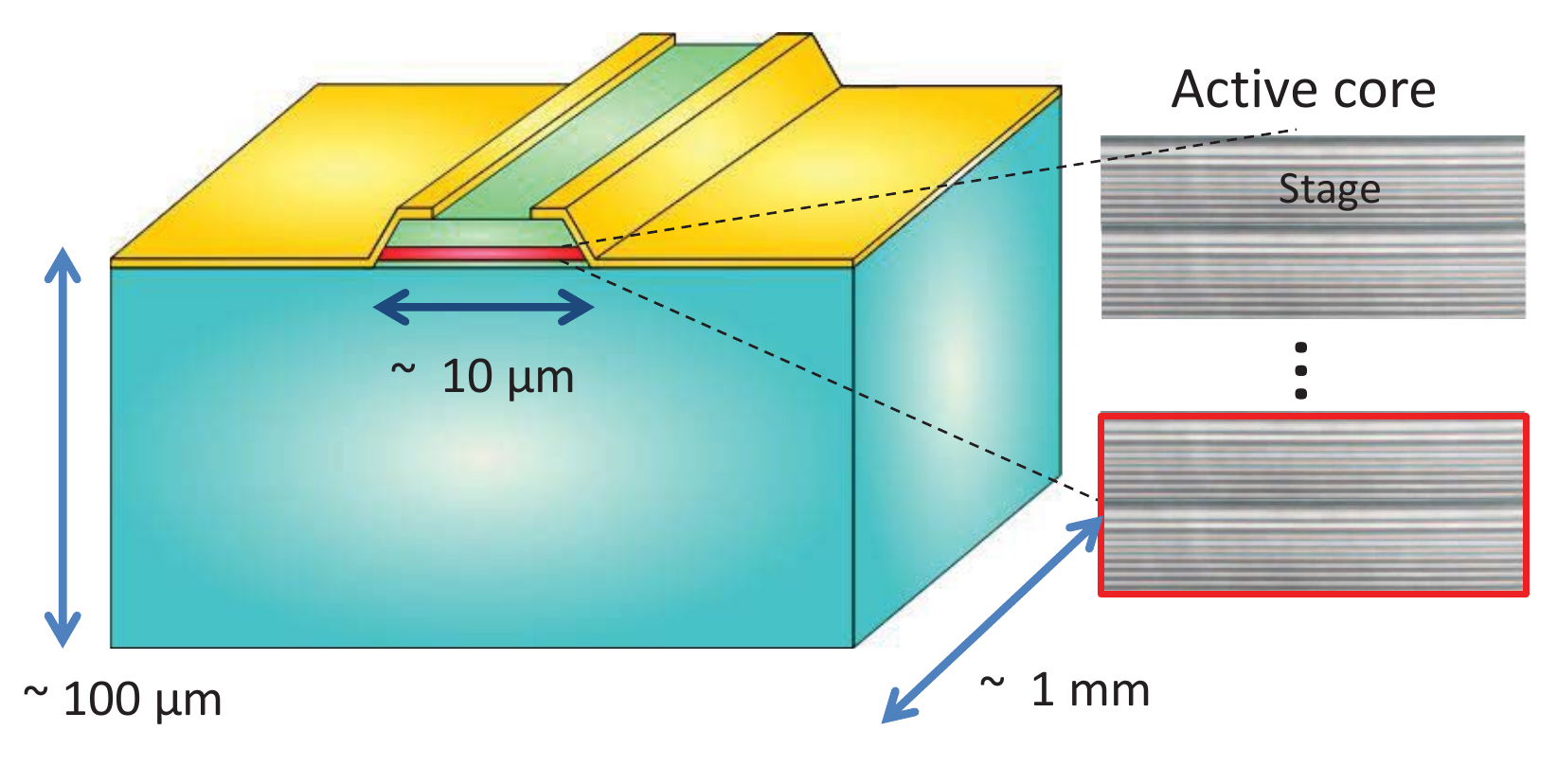}%
    \caption{\label{fig:multiscale}
    The multiscale nature of the QCL transport problem. While electron   transport and optical-field emission occur in the active core of the device and can be electrically controlled, thermal transport involves the entire large device and is only controlled via thermal boundary conditions that can be far from the active core. }
\end{figure}

Anharmonic decay is typically an order of magnitude slower than the rate at which the electron system deposits energy into the LO-phonon system. The fast relaxation of electrons into LO phonons, followed by the LO phonon slower decay into LA phonons, results in excess nonequilibrium LO phonons that can have appreciable feedback on electronic transport, population inversion, and the QCL figures of merit. As a result, different stages in the active core will have temperatures different from one another and drastically different from the heat sink (see Fig. \ref{fig:multiscale}). The electronic temperatures are higher still, differing among subbands, and affecting leakage paths and thus QCL performance \cite{Jonasson2016Photonics,botez2013multidimensional,botez2015temperature,bugajski2014mid}. In order to accurately describe QCL performance in the far-from-equilibrium conditions of CW operation, a multiscale electrothermal simulation is needed.

In this paper, we overview our recent work on developing a simulation framework capable of capturing the highly nonequilibrium physics of the strongly coupled electron and phonon systems in QCLs. In mid-IR devices \cite{Wacker2002a,JirauschekKubisAPR2014}, both electronic and optical phonon systems are largely semiclassical and described by coupled Boltzmann transport equations, which we solve using an efficient stochastic technique known as ensemble Monte Carlo \cite{Jacoboni1983}. The optical phonon system is strongly coupled to acoustic phonons, the dominant carreirs of heat, whose dynamics and thermal transport throughout the whole device are described via a global heat-diffusion solver. We discuss the roles of nonequilibrium optical phonons in QCLs at the level of a single stage \cite{Shi2014}, anisotropic thermal transport of acoustic phonons in QCLs \cite{Mei2015}, outline the algorithm for multiscale electrothermal simulation, and present preliminary data for a mid-IR QCL based on this framework.

\section{Single-stage simulation: Solving coupled electron and optical-phonon Boltzmann equations}\label{sec:single stage}

The model QCL is a 9-$\mu{{m}}$ QCL of Page \textit{et al.} \cite{Page2001}. Confined electronic states in every stage were calculated based on the envelope function approximation and $\mathbf{k\cdot p}$ formalism, known to be accurate in compound semiconductros near the $\Gamma$ point \cite{Gao2006,Gao2007,Gao2008,Shi2014}. In QCLs with appreciable doping, band bending is important, and we solve the coupled Schr\"{o}dinger equation (the envelope function and $\mathbf{k\cdot p}$) and Poisson's equation to obtain the accurate band profile \cite{Gao2007}. We assume bulk phonons, as it was shown that phonon confinement has little effect on electronic transport in mid-IR QCLs \cite{Gao2008}.

\begin{figure}[h!]
\centering
  \includegraphics[width=2.5 in]{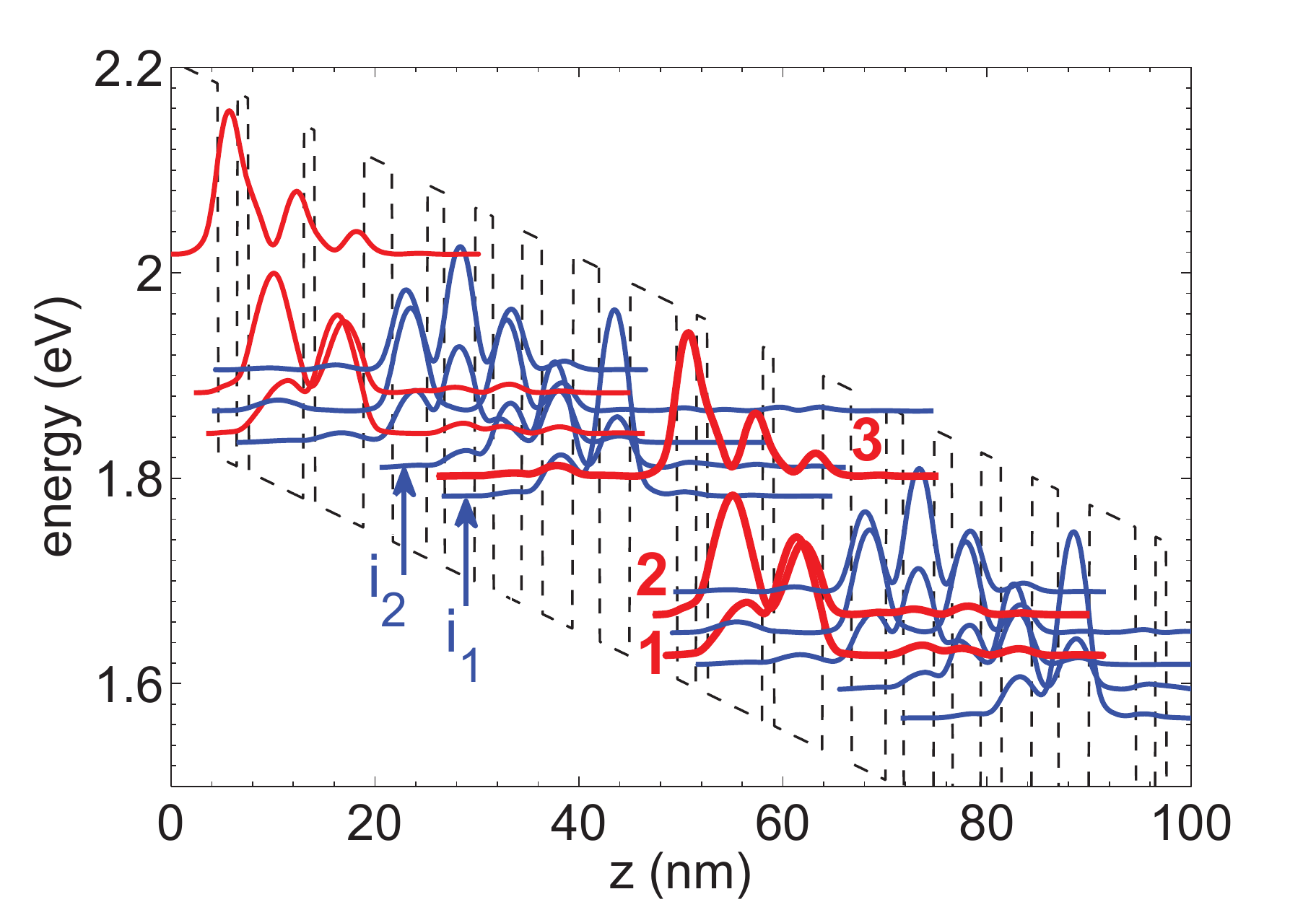}%
  \caption{\label{fig:bandstructure}
   Energy levels and wave-function moduli squared of $\Gamma$-valley subbands in two adjacent stages of the simulated GaAs/AlGaAs-based structure used in this preliminary work \cite{Page2001}. The bold red curves denote the active-region states (1, 2, and 3 represent the ground state and the lower and upper lasing levels, respectively). The blue curves represent injector states, with $i_1$ and $i_2$ denoting the lowest two. Reproduced from \cite{Shi2014}, Y. B. Shi and I. Knezevic, J. Appl. Phys. 116, 123105 (2014), with the permission of AIP Publishing.}
\end{figure}

\subsection{Electron and phonon Boltzmann transport equations} 
\begin{figure}
  \centering
  \includegraphics[width=0.6\columnwidth]{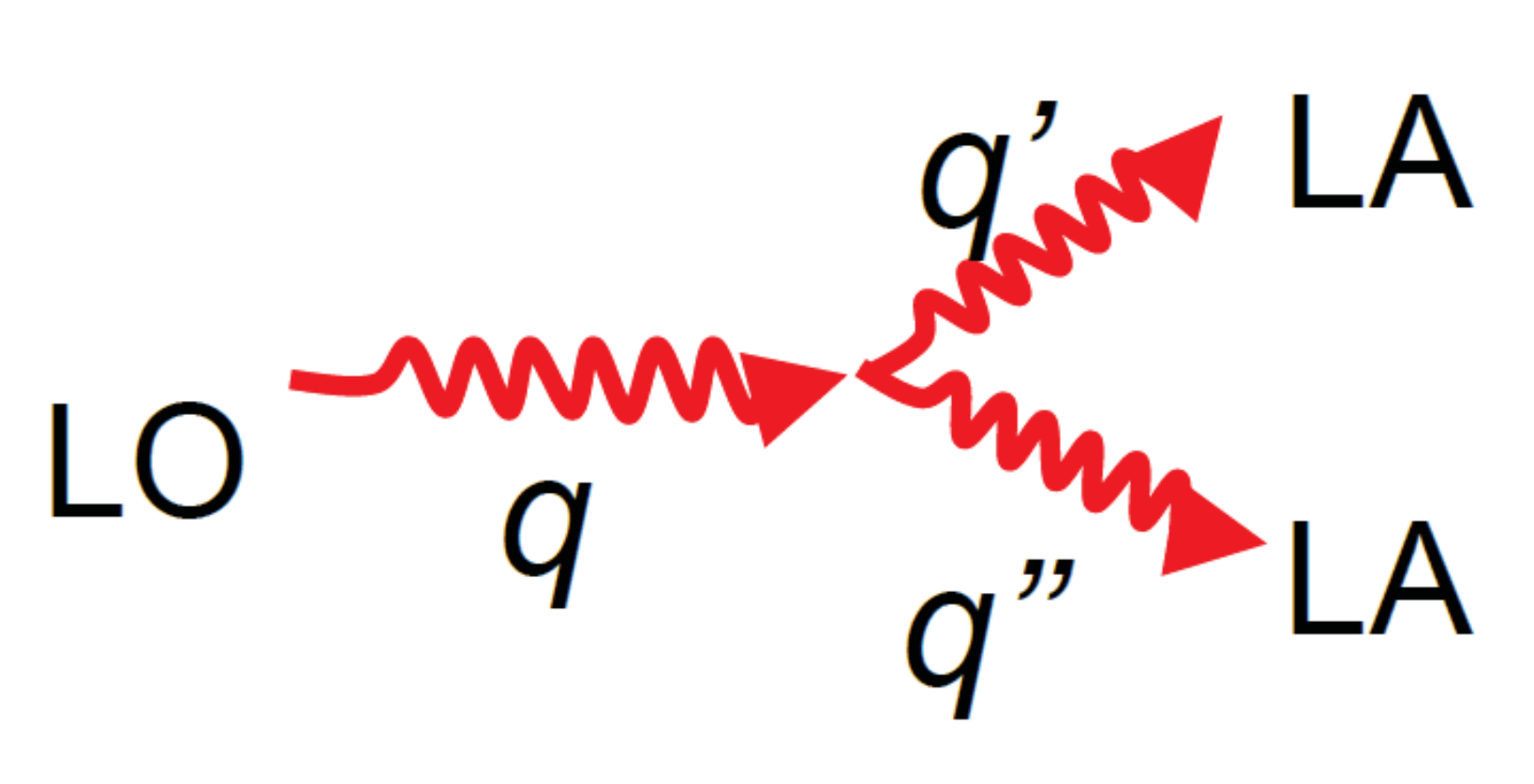}%
  \caption{\label{fig:anharmonic}
    The main anharmonic decay channel in III-V QCLs: decay of a longitudinal optical (LO) phonon into two longitudinal acoustic (LA) phonons.}
\end{figure}

In III-V semiconductors, electrons couple strongly to optical phonons. The materials are polar and optical phonons lead to electron scattering from the accompanying dipole field; we usually speak of emission or absorption of polar optical phonons by electrons. The coupling is strongest with LO phonons, with room-temperature emission rates on the order of $10^{12}$--$10^{13}\,{s}^{-1}$ (i.e., electron lifetimes on the order of 0.1 -- 1 ps). LO phonons decay into LA phonons about an order of magnitude more slowly (rate of $10^{11}\,{s}^{-1}$, i.e., LO-phonon lifetime of about  10 ps for the anharmonic decay into two LA phonons of different energies; see Fig. \ref{fig:anharmonic}). We solve the coupled electron and phonon Boltzmann transport equations, with the net phonon generation rate (emission minus absorption rate) to couple them \cite{Shi2014}. The phonon equation has both a net generation term from electrons and the anharmonic decay term; the latter yields the heat-generation rate $Q$ for the global heat diffusion solver \cite{Shi2012}.

\begin{figure}
  \includegraphics[width=\columnwidth]{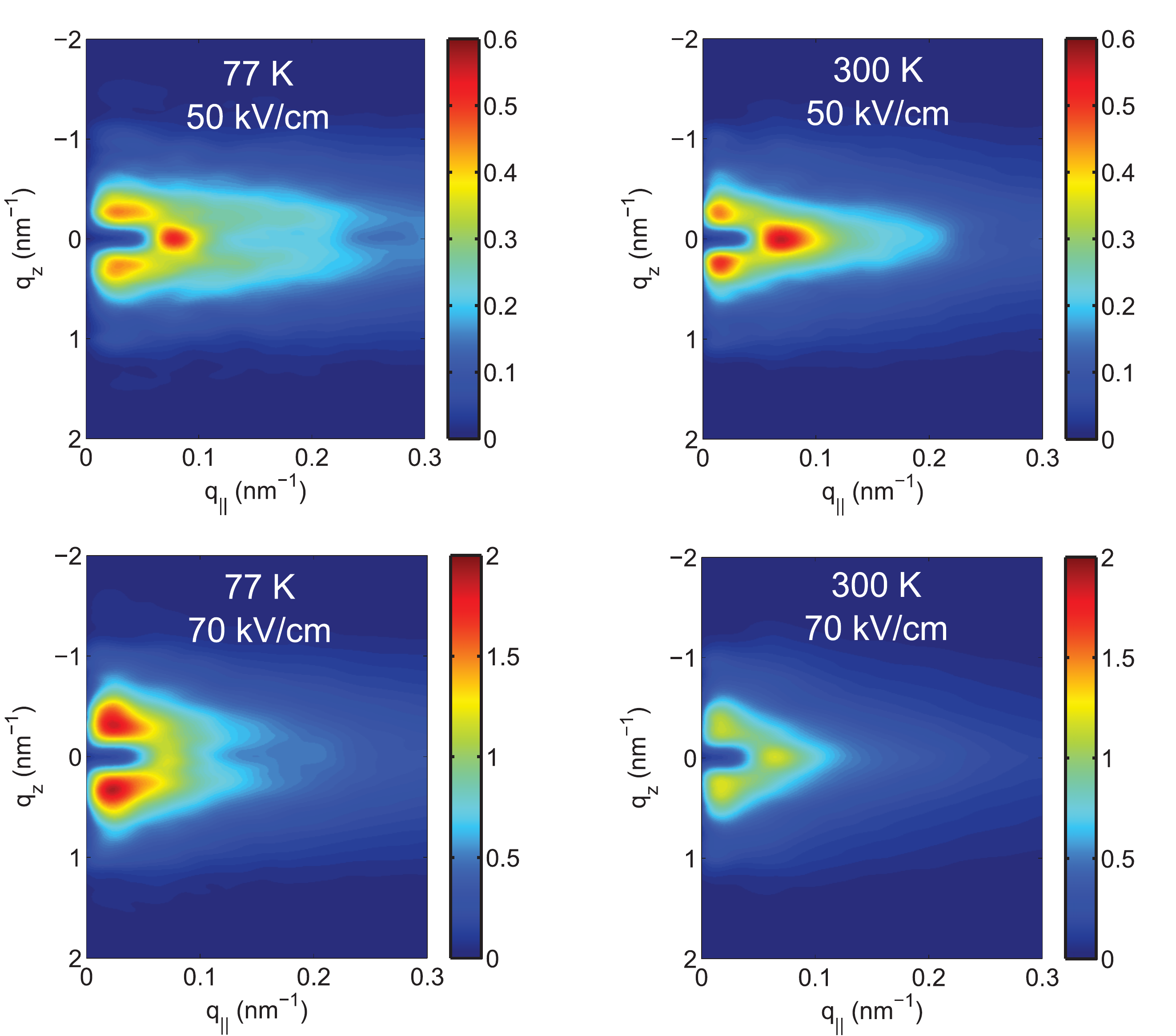}%
  \caption{\label{fig:noneqphonons}
The occupation number of excess nonequilibrium LO phonons, $N_\mathbf{q}-N_0$, versus the magnitude of the radial in-plane momentum $q_{||}$ and the cross-plane momentum $q_z$, presented via color (red--high, blue--low) at temperatures of 77 K and 300 K and fields of 50 kV/cm and 70 kV/cm in a 9 $\mu$m GaAs-based QCL \cite{Page2001}.  Note the different color bars that correspond to different fields.  Reproduced from \cite{Shi2014}, Y. B. Shi and I. Knezevic, J. Appl. Phys. 116, 123105 (2014), with the permission of AIP Publishing.}
\end{figure}

To solve these coupled Boltzmann equations, we employ the stochastic ensemble Monte Carlo technique \cite{JirauschekKubisAPR2014,Iotti2001,Gao2006,Gao2007}. A large ensemble of numerical electrons is tracked over time as they undergo periods of free flight in plane, which are interrupted by scattering events that may result in electrons changing their subband and/or in-plane state. The scattering processes we consider are electron--LO-phonon scattering and electron--electron scattering  \cite{Lundstrom92,JirauschekKubisAPR2014}. (Interface roughness is important in the optical response \cite{JirauschekKubisAPR2014}; it is at present not included in the electron transport simulation, but is a planned near-future addition.) Solving the BTE for electrons is well justified in mid-IR QCLs, where it has been shown that electronic transport is largely incoherent \cite{JirauschekKubisAPR2014}. Namely, in the basis of energy eigenstates associated with stages \cite{Wacker2002a}, which themselves do not carry current as they are localized, calculating the current density requires calculating off-diagonal density matrix elements (coherences) between these localized states. When these coherences are very small, they can be approximated as proportional to the occupations of states times the rates of scattering out of the states, which yields the hopping semiclassical picture of transport \cite{Lee2006QCL}. This picture fails in the limit of infrequent scattering and for large coherences, in general, such as in THz QCLs \cite{Jonasson2016}.

The phonon Boltzmann transport equation produces a phonon histogram, recording the phonons that are created and those that are destroyed either by re-absorption by electrons (not necessarily through the same process that created them) or due to the decay into acoustic phonons  (considered a random process;  the anharmonic decay time and random numbers are used to simulate the LO phonons decaying into LA ones). Phonon-phonon scattering rates for many III-Vs are known, the rest can be obtained from first principles, using density-functional perturbation theory (DFPT) \cite{Broido2007}. This histogram is also used to recalculate, ``on the fly'' \cite{Shi2014}, the electron--phonon scattering rates, which are proportional to phonon occupation numbers. Thereby, this simulation self-consistently accounts for electron scattering via nonequilibrim LO phonons.

\begin{figure}
  \includegraphics[width=\columnwidth]{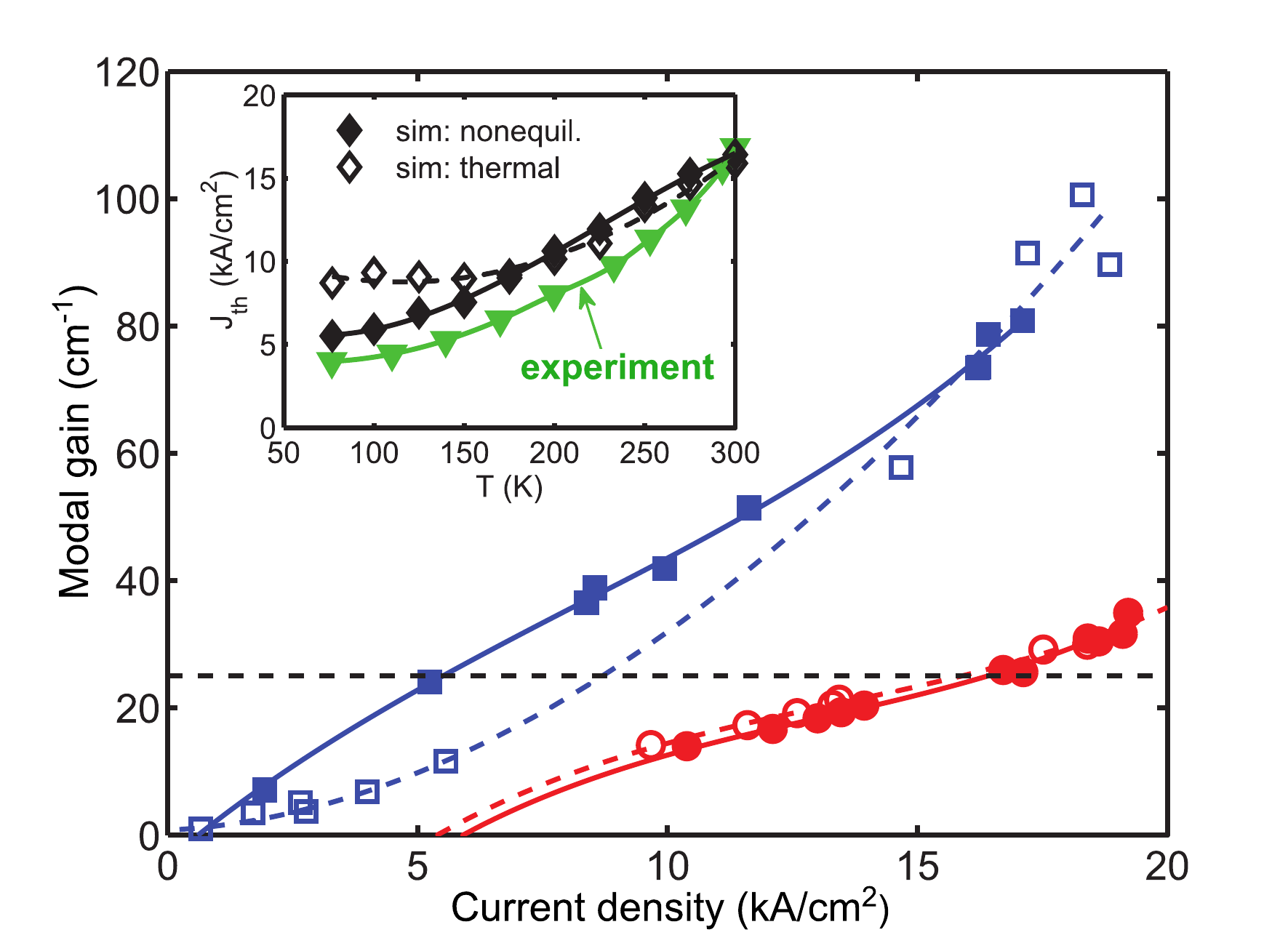}%
  \caption{\label{fig:modalgain}
Modal gain as a function of current density, obtained from the simulations with nonequilibrium (solid curves) and thermal (dashed curves) phonons
at 77 K (squares) and 300 K (circles). The horizontal dashed line denotes the estimated total loss of about 25 ${cm}^{-1}$. Inset: Threshold current density vs lattice temperature, as calculated with nonequilibrium (solid diamonds) and thermal (open diamonds) phonons, and as obtained from experiment \cite{Page2001} (triangles). Reproduced from \cite{Shi2014}, Y. B. Shi and I. Knezevic, J. Appl. Phys. 116, 123105 (2014), with the permission of AIP Publishing.}
\end{figure}

Optical phonons in GaAs and other III-V's have energies in the tens of meV range, so their thermal occupation number, $N_0$, is low even at room temperature. For instance, in GaAs,  the optical phonon energy is 36 meV, so $N_0(77 K)=6\times 10^{-4}$, $N_0(300 K)=0.33$. However, in high electric fields, considerable populations of nonequilibrium optical phonons are achieved. In Fig. \ref{fig:noneqphonons}, we see the occupation number (color) of excess nonequilibrium optical phonons, $N_\mathbf{q}-N_0$, versus phonon cross-plane momentum $q_z$ and in-plane momentum $q_{||}$. In the high fields (70 kV/cm), occupations of zone-center phonons  can get as high as 2, nearly four orders of magnitude higher than the thermal phonon occupation at 77 K.
This drastic enhancement in phonon population affects electronic transport, especially at low temperatures. It chiefly amplifies electron absorption of phonons: the rate of absorption is proportional to $N_\mathbf{q}$, so it increases orders of magnitude due to nonequilibrium phonons at 77 K; the effect is much weaker on phonon emission, whose rate is proportional to $N_\mathbf{q}+1$. The increased absorption is actually beneficial, because the lowest injector level is typically below the upper lasing level in energy and phonon absorption is required for injection; the fact that nonequilibrium phonons drastically amplify the rate of phonon absorption means that the injection rate from the lowest injector level into the upper lasing level is enhanced, which helps with the population inversion and modal gain at low temperatures. Essentially, at low temperatures, where the nonequilibrium phonon number is considerable with respect to thermal phonon number, nonequilibrium phonons improve injection efficiency and modal gain at a given field, and thus reduce the threshold current density. Our work \cite{Shi2014} was the first to show how the presence of nonequilibrium phonons affects the QCL performance metrics that are accessed in experiment, such as the $J-F$ curves, modal gain (Fig. \ref{fig:modalgain}), and threshold current vs heat-sink temperature (inset to Fig. \ref{fig:modalgain}). We note that the inclusion of nonequilibrium phonons is very important for getting close to the experimentally observed shape of $J_{th}$ vs $T$; simulation with nonequilibrium LO phonons gives a much more accurate account of this dependence than the simulation with thermal phonons. Note the lowering of the threshold current density at low temperatures in the inset to Fig. \ref{fig:modalgain} because of nonequilibrium phonons.

\section{Thermal conductivity tensor in a QCL device}\label{sec:thermal cond QCL}

\begin{figure}
  \includegraphics[width=\columnwidth]{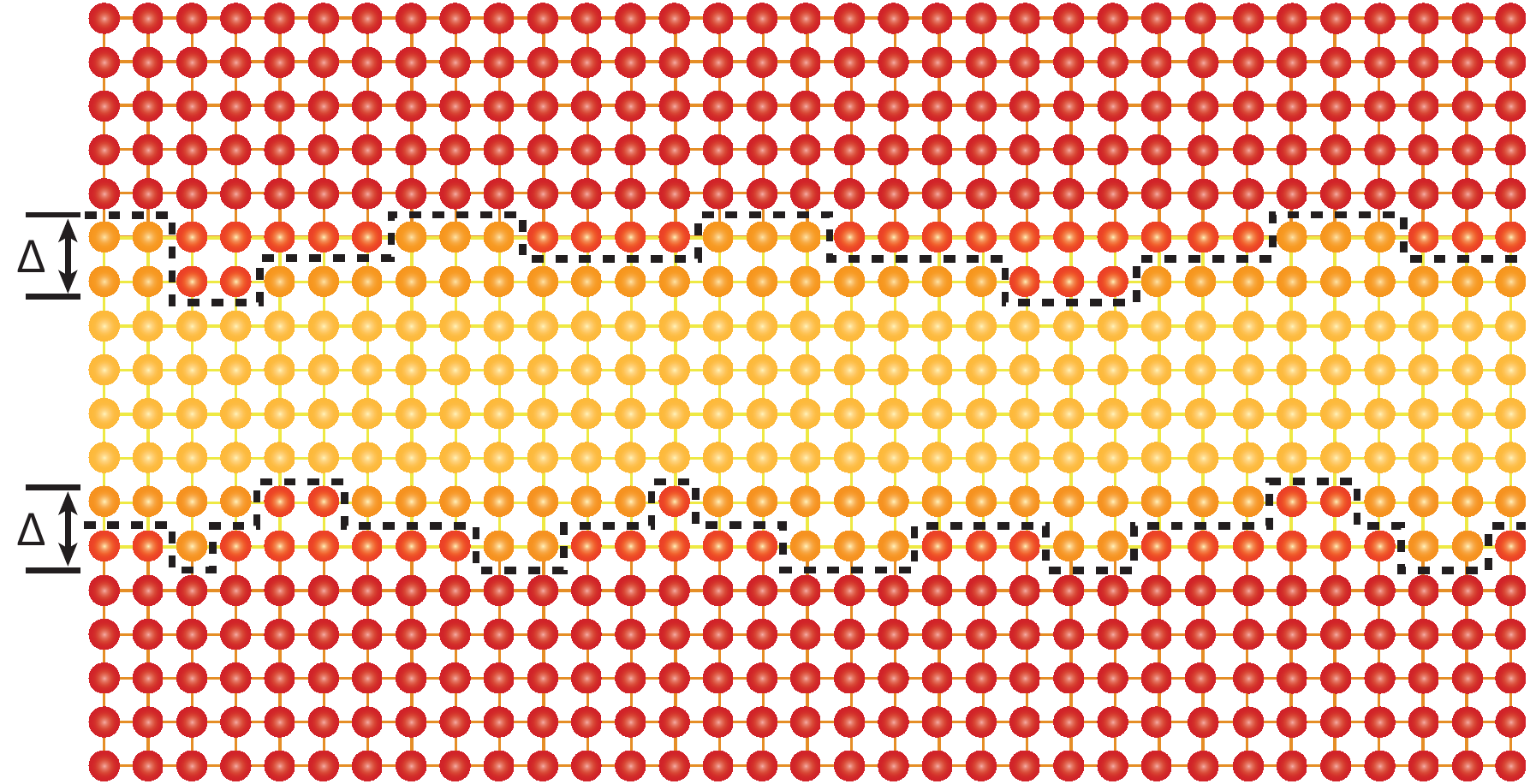}
  \caption{\label{fig:superlatticeschematic}
Even between lattice-matched crystalline materials, there exist nonuniform
transition layers that behave as an effective atomic-scale interface
roughness with some rms height $\Delta$. This effective interface roughness leads to phonon-momentum randomization and to interface resistance in crossplane transport. Reproduced from \cite{Mei2015}, S. Mei and I. Knezevic, J. Appl. Phys. 118, 175101 (2015), with the permission of AIP Publishing.}
\end{figure}

In most nanostructures and at temperatures that are not too low, phonon transport (dominated by acoustic phonons) is diffusive and obeys the Boltzmann transport equation. Heterostructures are known to be extremely anisotropic thermal conductors, with low cross-plane and high in-plane thermal conductivity \cite{Cahill2003,Cahill2014,Chen1998,Vitiello2008}. The anisotropy in thermal conductivity largely stems from anisotropic dispersions (different phonon group velocities in different directions) \cite{Simkin2000} and transmission through the interface (Fig. \ref{fig:superlatticeschematic}), which may be accompanied by some randomization of phonon momentum due to atomic-scale interface roughness \cite{Aksamija2013,Cahill2014,Chen1997,Chen1998}. To calculate the thermal conductivity of III-V QCLs, we solved the Boltzmann transport equation for phonons in the relaxation-time approximation \cite{Mei2015}.

In essence, the influence of the interface roughness is twofold. First, it lowers the in-plane layer thermal conductivity by affecting the population of acoustic phonon modes. Second, there is interface boundary resistance which is notoriously difficult to understand and compute. It comprises the effects of acoustic mismatch (analogue of Snell's law, with different materials having different acoustic impedances) and diffuse scattering at a momentum-randomizing interface. Indeed, the two limiting models for interface boundary resistance stem from the acoustic mismatch model (AMM) and the diffuse mismatch model (DMM), neither of which doing justice to the high quality interfaces typical for III-V heterostructures. We introduced a simple model that  interpolates between the two, with the weight of the AMM transmission coefficient in the interpolation being the same momentum-dependent specularity parameter related to roughness that is routinely used to address slightly rough nanostructures. We note that the concept of the specularity parameter assumes that there is no lateral correlation \cite{Soffer1967,ZimanBook,Mei2015}. Increasing the correlation length makes an interface appear more specular \cite{Soffer1967,Maurer2015}, so interface correlation can be captured in part with a lower effective rms roughness and a higher effective specularity parameter \cite{Mei2015}. We were able to fit a large number of disparate measurements by different groups on thermal conductivities of binaries, ternaries, and superlattices. This technique for calculating the thermal conductivity tensor can also readily account for strain, which is important in short-wavelength QCLs.

\begin{figure}
  \includegraphics[width=\columnwidth]{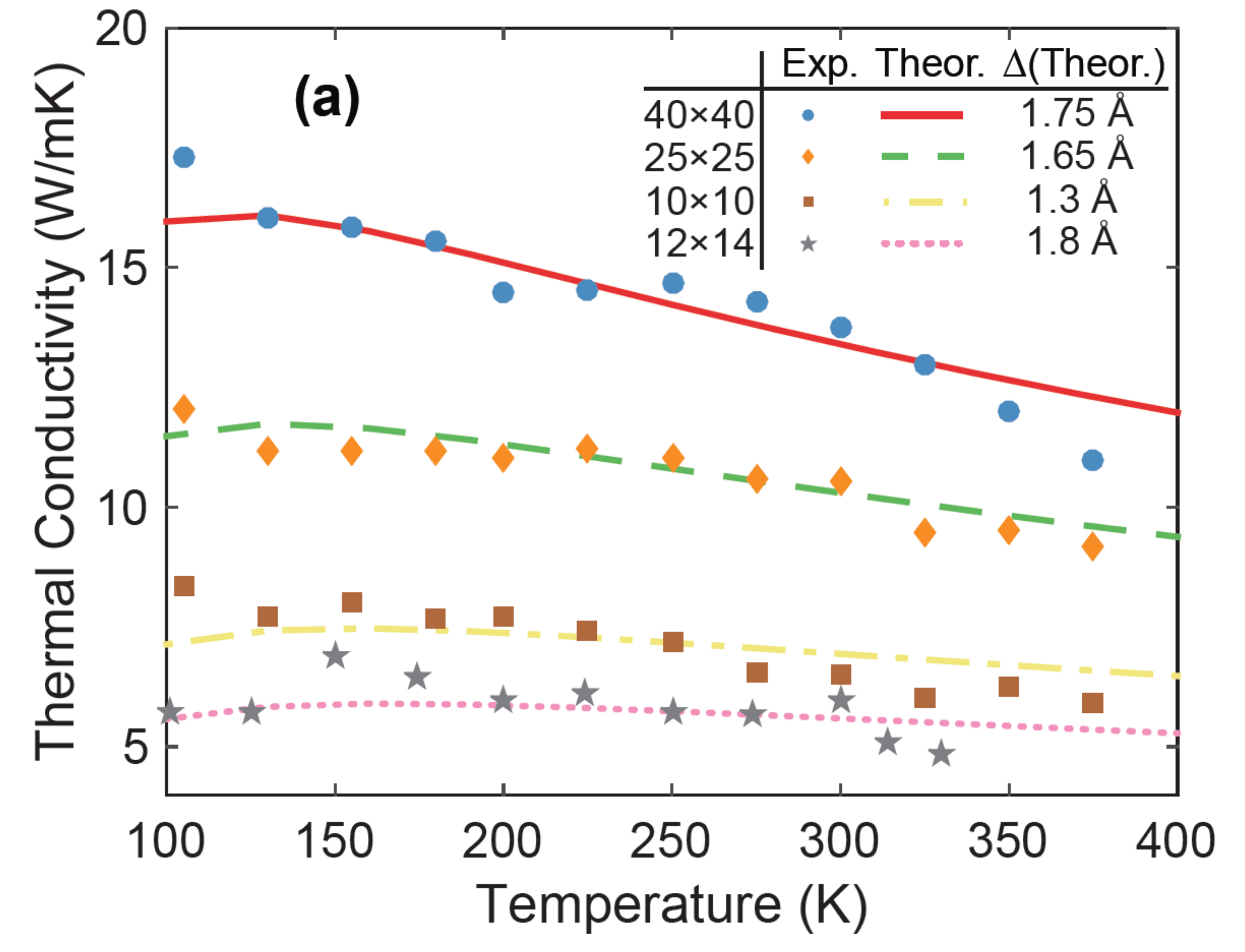}
  \caption{\label{fig:thermalcondsuperlatt}
Cross-plane thermal conductivity of GaAs/AlAs superlattices as
a function of temperature. Blue circles, orange diamonds, and brown squares
show the measured cross-plane thermal conductivity data for 40$\times$40,
25$\times$25, and 10$\times$10 superlattices from Ref. \cite{Capinski1999}. Grey stars are the cross-plane thermal
conductivity data for a 12$\times$14 superlattice from Ref. \cite{CAPINSKI1996}. The corresponding curves are calculated based on our model, with the optimal effective rms roughness $\Delta$ denoted in the legend. Reproduced from \cite{Mei2015}, S. Mei and I. Knezevic, J. Appl. Phys. 118, 175101 (2015), with the permission of AIP Publishing.}
\end{figure}

\begin{figure}
  \includegraphics[width=\columnwidth]{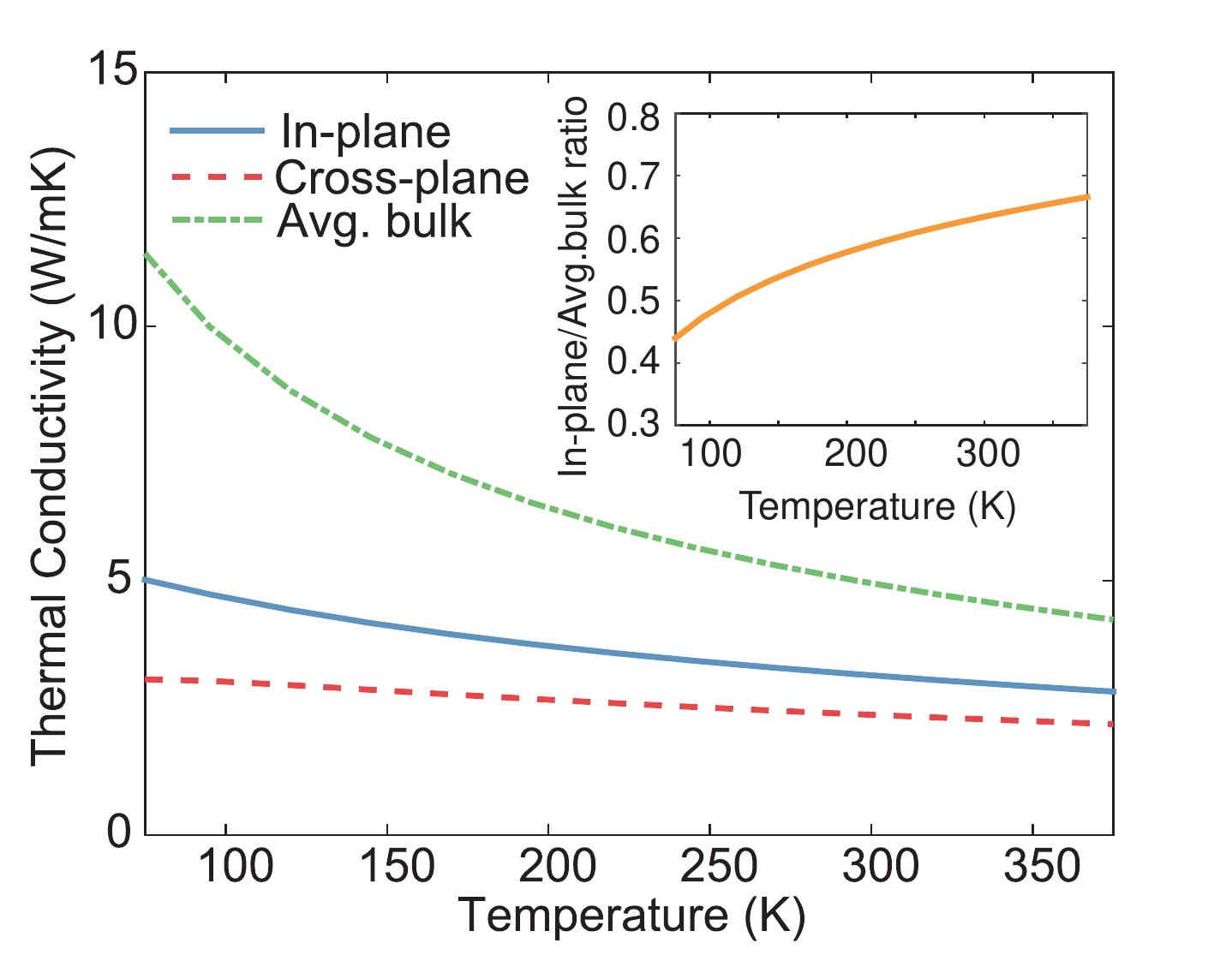}
  \caption{\label{fig:anisotropy}
Thermal conductivity of a typical QCL active region \cite{Lops2006} as a function of temperature. A single stage consists of 16 alternating layers of $In_{0.53}Ga_{0.47}As$ and $In_{0.52}Al_{0.48}As$. Blue solid curve, red dashed curve, and green dashed-dotted curve are showing the calculated in-plane, cross-plane, and the averaged bulk thermal conductivity, respectively. $\Delta$=1 {\AA} in the calculations. The inset shows the ratio between the calculated in-plane and the averaged bulk thermal conductivities. Reproduced from \cite{Mei2015}, S. Mei and I. Knezevic, J. Appl. Phys. 118, 175101 (2015), with the permission of AIP Publishing.}
\end{figure}

We calculated the thermal conductivity tensor in a number of III-V superlattices and QCL structures over a range of temperatures \cite{Mei2015}. In Fig. \ref{fig:thermalcondsuperlatt}, we show the calculated thermal conductivity of GaAs/AlAs superlattices from different experiments \cite{CAPINSKI1996,Capinski1999,Mei2015}. The results for GaAs and AlAs involve three-phonon (normal and umklapp) scattering for acoustic braches and isotope scattering \cite{Mei2015}. All the results incorporate full phonon dispersions. The experimental superlattices apparently have good-quality interfaces, with rms roughness of only 1-2 angstroms (obtained from fitting theory to experiment with the rms roughness as the only fitting parameter); the small rms roughness  is related to the wide, atomically flat terraces. We also note that the anisotropy (cross-plane vs in-plane ratio) is temperature dependent (Fig. \ref{fig:anisotropy}). The in-plane vs bulk ratio is also temperature dependent and is not the often-assumed value of 75\%; rather, it varies from 40 to 70\%, depending on temperature. This data underscores the importance of careful calculation of the thermal conductivity tensor.

\section{Device-level electrothermal simulation}\label{sec:multiscale}

\begin{figure}
  \centering
  \includegraphics[width=2.5 in]{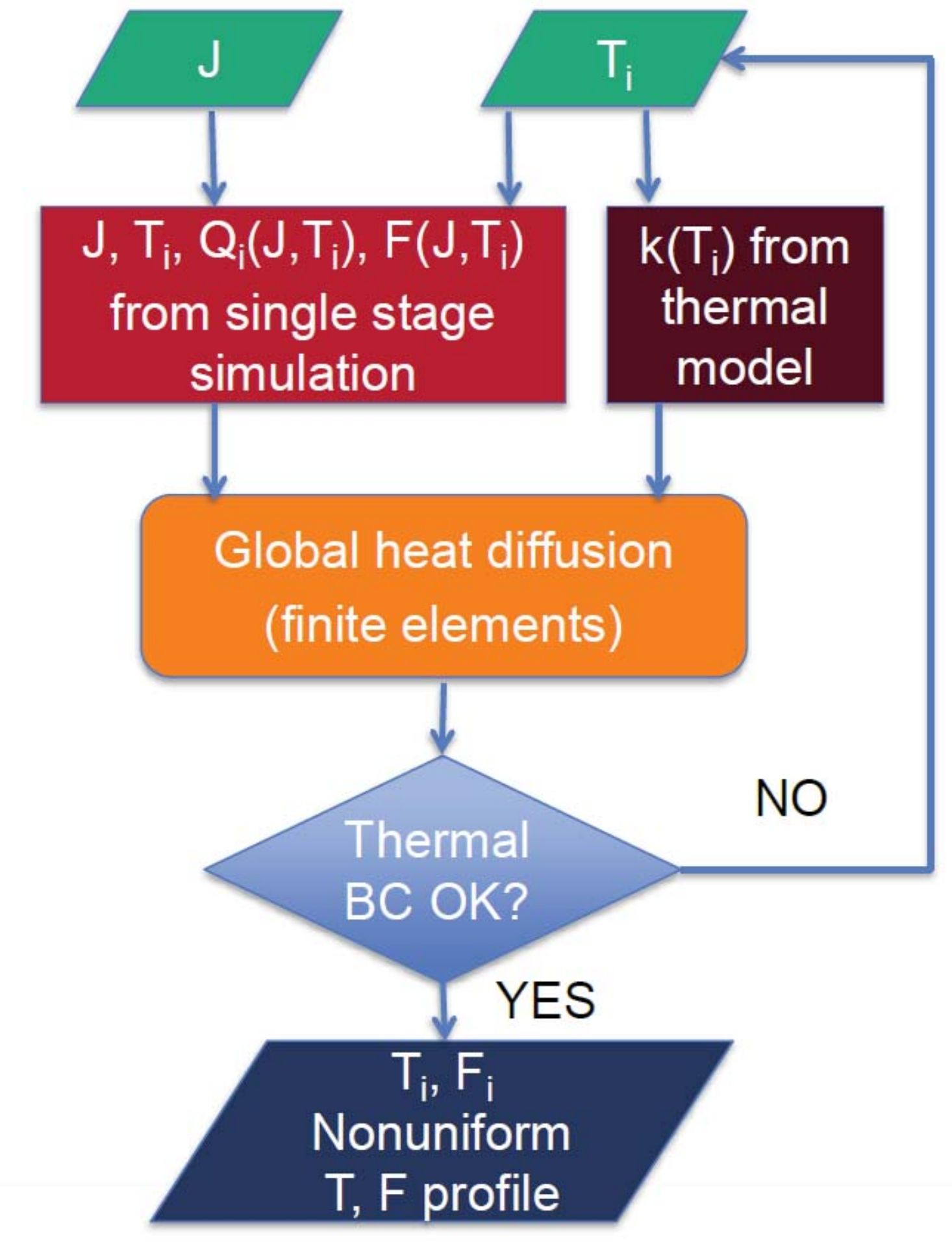}%
  \caption{\label{fig:globalsimulationflowchart}
    Flowchart of the global simulation. We start by assuming a certain current density ${J}$ and temperature profile ${T_i}$ across the whole device; for example, we can start by assuming the whole device is at the heat-sink temperature.  Based on the tabulated information from the single-stage simulation and assumed ${(J,T_i)}$, we get stage-by-stage profile for the electric field ${F_i}$ and the heat current density ${Q_i}$. We use the temperature profile guess and an accurate temperature-dependent thermal conductivity model, which includes the boundary resistances of layers, as input to the heat-diffusion equation. The heat-diffusion equation is solved iteratively, with updated temperature profile in each step, until the thermal boundary conditions are satisfied.}
\end{figure}

\textbf{Single-stage coupled simulation.} First, we develop a fully coupled electronic and thermal transport simulation for each stage; this is achieved by solving coupled Boltzmann transport equations for electrons and LO phonons via the stochastic ensemble Monte Carlo technique \cite{Jacoboni1983}. A stage is characterized by starting with an assumed average electric field $F$ for that stage and a lattice temperature $T_L$, the latter giving baseline phonon occupations and baseline electron--optical phonon scattering rates. For each field $F$ and lattice temperature $T_L$ (assumed to coincide with the acoustic-phonon ensemble temperature), the key output consists of the electrical current density $J$ and the heat-generation rate $Q$; $Q$ tells us about the rate at which acoustic phonons are generated by the decaying optical phonons. By sweeping $F$ and $T_L$, single-stage simulation of coupled electron and phonon transport yields a ``table" connecting different pairs of field and lattice temperature to the appropriate pairs of current density and heat-generation rate, i.e., we have a tabulated map $(T_L, F) \rightarrow (J, Q)$.

\textbf{Current continuity.} There is no guarantee that the field or temperature is the same in every stage of the active region. In fact, the field variation between stages is a well-known staple of superlattices, but for some reason underappreciated in QCLs \cite{Wacker2002a}. While we cannot assume stage-independent temperature or field, the charge--current continuity equation certainly holds. Therefore, in the steady state, the current density $J$ must be the same in every stage. So we ``flip'' the table obtained in the single-stage simulation, from $(T_L, F) \rightarrow (J, Q)$ to $(J,T_L) \rightarrow (Q,F)$. Therefore, input from single-stage simulation into global simulation is the flipped table $(J,T_L) \rightarrow (Q,F)$.

\textbf{Heat flow through the whole device.} In essence, the global simulation is the solution to the heat-diffusion equation, with each stage acting as a current-dependent heat source. However, we do not know what temperature each stage might have; all we can assume are a current density $J$ and certain thermal boundary conditions (heat-sink temperatures or convection boundary conditions). We can in principle calculate the temperature-dependent thermal conductivity tensor everywhere in the device (see Sec. \ref{sec:thermal cond QCL}). Therefore, the input for global simulation is the thermal conductivity tensor $\kappa (T)$ at every point in the whole device.

\textbf{Putting it all together.} Based on single-stage simulation, we have created a large table of $(J,T_L) \rightarrow (Q,F)$ maps. Then, for a given current density level $J$ and a given set of thermal boundary conditions (heat-sink temperatures or convection boundary conditions at exposed facets), we assume a temperature profile throughout the device. In each stage $i$, the $J$ and a  guess for the stage temperature $T_i$ produce the appropriate heat-generation rate in that stage, $Q_i(J,T_i)$. The guess temperature profile $T_i$, heat-generation rate $Q_i$, and the thermal model yielding the thermal conductivity tensor everywhere are used in the heat-diffusion equation  (which is solved via the standard finite elements technique) to calculate the updated temperature profile. The process is iterated until the obtained temperature profile agrees well with the imposed thermal boundary conditions. Upon the calculation of the temperature profile, we can read off the stage field ${F_i(J,T_i)}$ and calculate the actual complete voltage drop across the device and total optical power, which then go into calculating the QCL figures of merit.


\subsection{Example: Nonuniform field and temperature inside the active core}\label{sec:example}

\begin{figure}
  \includegraphics[width=\columnwidth]{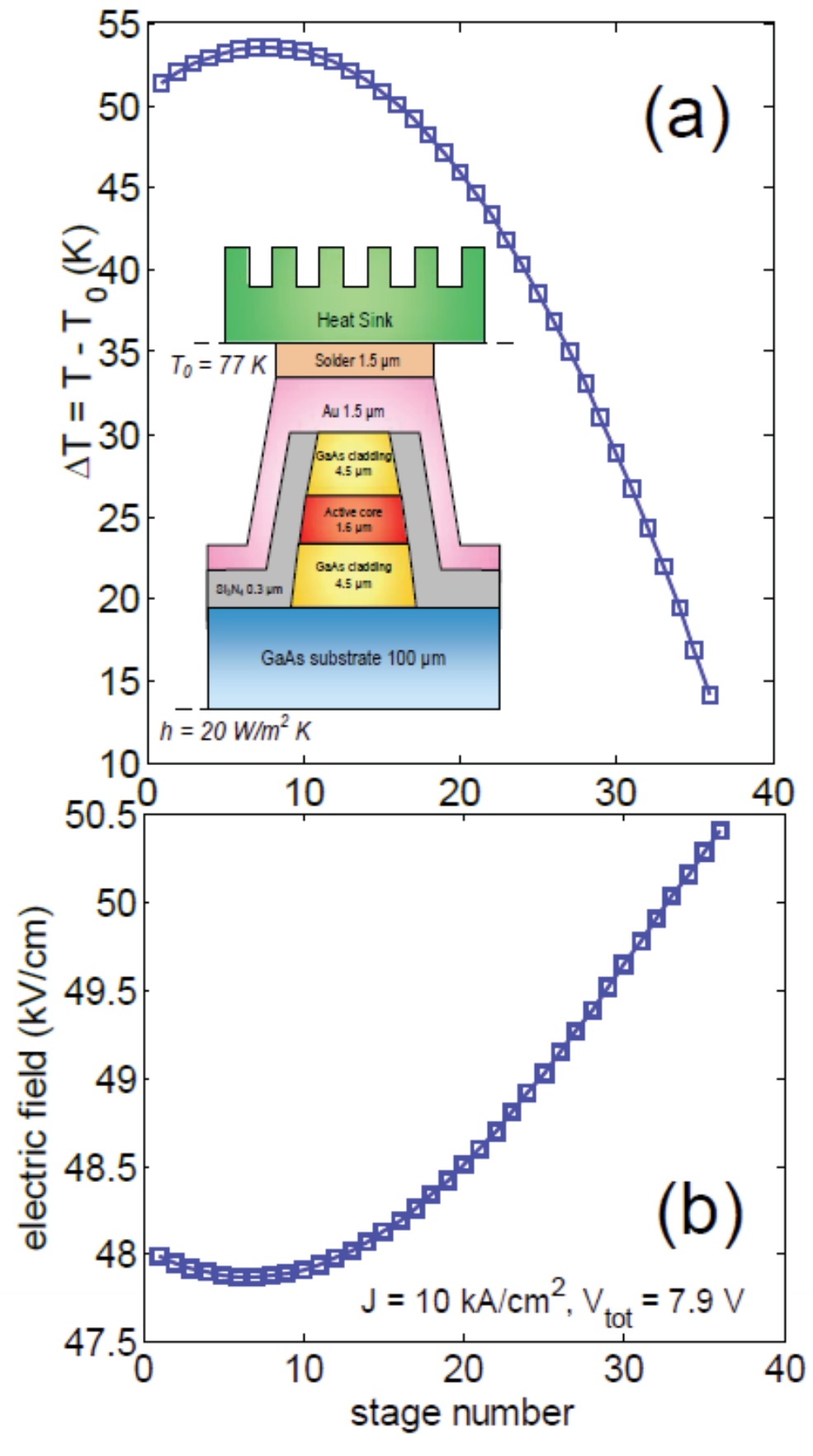}%
  \caption{\label{fig:nonuniformprofiles} (a) ${\Delta T}$, the difference between the lattice temperature in a given stage of the active core and the heat sink for the 9-${\mu m}$ GaAs device of Page \textit{et al.} \cite{Page2001}, simulated with our global algorithm. (Inset) Schematic of the device. The two boundaries of the active core, clearly visible in the schematic, correspond to stage $i=0$ from the main panel at the substrate end (bottom) and to stage $i=40$ from the main panel close to the heat sink (top). The heat-sink temperature is 77 K, convection boundary condition is assumed at the substrate end, and the current density is ${J=10\,{kA/cm^2}}$. (b) The electric-field profile in the active core at the same current density ${J}$.}
\end{figure}

This data was obtained based on a preliminary implementation of the device-level (global)  simulation described above. In Fig. \ref{fig:nonuniformprofiles}, we present the results for the electric field and temperature profile inside the active core at a current density of $10\, kA/cm^2$. First, the temperature in the active core is quite high, being from 10 to over 50 degrees above the heat-sink temperature of 77 K [Fig. \ref{fig:nonuniformprofiles}(a)]. Such high temperatures may lead to considerable thermal stress between layers and between the active core and the substrate, which may lead to material damage after prolonged use \cite{Zhang2010}.  	
The nonuniform field profile [Fig. \ref{fig:nonuniformprofiles}(b)] implies that growing all stages in the exact same way, for the same exact operating field, is not ideal because they cannot all be at the same field simultaneously. Identical stages will not all be working optimally, the line will be broadened and the optical power lowered. Instead,  each stage could ideally be optimized separately to have a slightly different operating field, in keeping with Fig. \ref{fig:nonuniformprofiles}(b).

\section{Conclusion}
Quantum cascade lasers are excellent for studying far-from-equilibrium transport: they feature strongly coupled electron and phonon systems and they are experimentally well characterized owing to their practical importance, so theoretical models can be readily tested. In this paper, we presented our ongoing work on multiscale electrothermal simulation of quantum cascade lasers on the example of an older, well-characterized device, the 9-micron GaAs-based QCL \cite{Page2001}. We discussed the single-stage coupled dynamics modeling \cite{Shi2014}, thermal transport characterization \cite{Mei2015}, outlined a new algorithm for multiscale simulation, and showcased preliminary results for nonuniform temperature and field profiles inside the device, which stem from fully coupled multiscale simulation and underscore the importance of global (device-level) analysis of QCLs.

\section*{Acknowledgement}
The authors gratefully acknowledge support by the U.S. Department of Energy (Basic Energy Sciences, Division of Materials Sciences and Engineering, Physical Behavior of Materials Program) Award No. DE-SC0008712. The work was performed using the resources of the UW-Madison Center for High Throughput Computing (CHTC).



\providecommand{\WileyBibTextsc}{}
\let\textsc\WileyBibTextsc
\providecommand{\othercit}{}
\providecommand{\jr}[1]{#1}
\providecommand{\etal}{~et~al.}

\end{document}